\documentclass[showpacs]{revtex4}
\usepackage{epsfig}
\def\beq{\begin{equation}}
\def\enq{\end{equation}}
\def\beqa{\begin{eqnarray}}
\def\enqa{\end{eqnarray}}
\def\MeV{\nobreak\,\mbox{MeV}}
\def\GeV{\nobreak\,\mbox{GeV}}
\def\keV{\nobreak\,\mbox{keV}}

\def\qq{\lag\bar{q}q\rag}
\def\ss{\lag\bar{s}s\rag}

\def\G3{\lag g^3G^3\rag}
\def\pli{p^\prime}

\def\la{\lambda}

\def\ga{\gamma}
\def\Ga{\Gamma}

\def\lb{\label}
\def\nn{\nonumber}
\newcommand{\rag}{\rangle}
\newcommand{\lag}{\langle}

\begin{document}

\title{\sc $D_{sJ}^+(2317)\to D_s^+\pi^0$ decay width}
\author{Marina Nielsen}
\affiliation{Instituto de F\'{\i}sica, Universidade de S\~{a}o Paulo, 
C.P. 66318, 05389-970 S\~{a}o Paulo, SP, Brazil}

\begin{abstract}
We use the QCD sum rules to
analize the hadronic decay $D_{sJ}^{+}(2317)\to D_s^+\pi^0$, in the 
hypotesis that the $D_{sJ}^{+}(2317)$ can be identified as a four-quark 
state. We use a diquak-antidiquark current and work to the order of $m_s$ 
in full QCD, without relying on $1/m_c$ expansion. We find that the
partial decay width of the hadronic isospin violating mode is proportional to
the isovector quark condensate, $\langle0|\bar{d}d-\bar{u}u|0\rangle$.
The estimated partial decay width is of the order of 6 keV.
\end{abstract}

\pacs{ 11.55.Hx, 12.38.Lg , 13.25.-k}
\maketitle

The strange-charmed mesons  $D_{sJ}^+(2317)$ and $D_{sJ}^+(2460)$ with spin 
parity $0^+$, $1^+$ \cite{babar,cleo,belle1,focus} are lighter than the
prediction of the very successful quark model for the charmed mesons 
\cite{god}. One interpretation is that this is evidence for a chiral
symmetry where the mass gap between the $0^-$ and $1^-$ states equates
the mass gap between the $0^+$ and $1^+$ states \cite{bar}. Other
interpretations are related with the underlying structure of these mesons,
which has been extensively debated. They have been interpreted as 
conventional $c\bar{s}$ states 
\cite{dhlz,bali,ukqcd,ht,nari,dnp,cf,god2,fr,whz}, 
two-meson molecular state \cite{bcl,szc}, $D-K$- mixing \cite{br} or 
four-quark states \cite{ch,tera,mppr,blmnn,kioh,bpp,dmi,hte}. 

Because of their low masses, these two states are lower than the $DK$ and
$D^*K$ thresholds. Therefore, their strong decays must proceed through isospin
violating effects. There have been some discussions of their decays within
the quark model \cite{bar,cf,god2,fr,ch} and QCD sum rules \cite{whz}. In
all these  studies but \cite{god2}, the isospin violating effects were 
considered through the $\eta-\pi^0$ mixing. However, if these mesons are 
considered as four-quark  states, in a QCD sum rule calculation, the isospin 
violating effects can be introduced through the mass and quark condensate 
difference between the $u$ and $d$ quarks.

In a recent calculation \cite{blmnn} the scalar-isoscalar meson 
$D_{sJ}^+(2317)$ were 
considered as a $S$-wave bound state of a diquark-antidiquark pair. As 
suggested in ref.~\cite{jawil}, the diquark was taken to be a spin zero 
colour anti-triplet. The corresponding interpolating field is: 
\beqa
j_S&=&{\epsilon_{abc}\epsilon_{dec}\over\sqrt{2}}\left[(u_a^TC
\gamma_5c_b)(\bar{u}_d\gamma_5C\bar{s}_e^T)+(d_a^TC
\gamma_5c_b)(\bar{d}_d\gamma_5C\bar{s}_e^T)\right],
\label{int}
\enqa
where $a,~b,~c,~...$ are colour indices and $C$ is the charge conjugation
matrix. In ref.~\cite{blmnn}, using the QCD sum rule (QCDSR) formalism
\cite{svz,rry,io1}, 
it was shown that it is possible to reproduce the experimental mass of the 
meson $D_{sJ}^+(2317)$ using this four-quark state picture. Here, we extend 
the calculation done in ref.~\cite{blmnn} to study the vertex associated with
the decay $D_{sJ}^+(2317)\to D_s^+\pi^0$. 

The QCDSR calculation for the vertex, $D_{sJ}^+(2317) D_s^+\pi^0$, centers
around the three-point function given by
\beq
\Gamma_\mu(p,\pli,q)=\int d^4x d^4y ~e^{i\pli.x}~e^{iq.y}
\lag 0 |T[j_{D_s}(x)j_{5\mu}^{\pi^0}(y)j^\dagger_S(0)]|0\rag,
\lb{3po}
\enq
where $p=\pli+q$ and the interpolating fields for the pion and $D_s$ mesons 
are given by:
\beq
j_{5\mu}^{\pi^0}={1\over\sqrt{2}}(\bar{u}_a\gamma_\mu\gamma_5u_a-\bar{d}_a
\gamma_\mu\gamma_5d_a),\,\;\;\;j_{D_s}=i\bar{s}_a\gamma_5c_a.
\lb{pseu}
\enq

The fundamental assumption of the QCD sum rule approach is the principle
of duality. Specifically, we assume that there is an interval over which
the above vertex function may be equivalently described at both, the quark 
level and at the hadron level. Therefore, the underlying procedure of the 
QCD sum rule technique is the following: on one hand we calculate the 
vertex function at the quark level in terms of quark and gluon fields. 
On the other hand, the vertex function is calculated at the hadronic level
introducing hadron characteristics such as masses and coupling constants.
At the quark level the complex structure
of the QCD vacuum leads us to employ the Wilson's operator product 
expansion (OPE). The calculation of the phenomenological side proceeds by 
inserting intermediate states for $D_s$, $\pi^0$ and $D_{sJ}$, and by using 
the definitions: 
\beq
\lag 0 | j_{5\mu}^{\pi^0}|\pi^0(q)\rag =iq_\mu F_{\pi},\;\;\;\;
\lag 0 | j_{D_s}|D_s(\pli)\rag ={m_{D_s}^2f_{D_s}\over m_c+m_s},\;\;\;\;
\lag 0 | j_{D_{sJ}}|D_{sJ}(p)\rag =\la_S.
\lb{fp}
\enq
We obtain the following relation:
\beqa
&&\Ga_{\mu}^{phen} (p,\pli,q)={\la_S m_{D_s}^2f_{D_s}F_{\pi}~
g_{D_{sJ}D_s\pi}
\over(m_c+m_s) (p^2-m_{D_{sJ}}^2)({\pli}^2-m_{D_s}^2)(q^2-m_\pi^2)}~
~q_\mu +\mbox{continuum contribution}\;,
\lb{phen}
\enqa
where the coupling constant $g_{D_{sJ}D_s\pi}$ is defined by the on-mass-shell
matrix element
\beq
\lag D_s \pi|D_{sJ}\rag=g_{D_{sJ}D_s\pi}.
\label{coup}
\enq
The continuum contribution in Eq.(\ref{phen}) contains the contributions of
all possible excited states.

For the light scalar mesons, considered as diquark-antidiquark states, the 
study of their vertices functions using the QCD sum rule approach at the pion 
pole \cite{rry,nari,nari2,brann}, was done in ref.\cite{sca}.
In Table I we show the results obtained for the different
vertices studied in ref.~\cite{sca}, as well as the experimental values.
\begin{center}
\small{{\bf Table I:} Numerical results for the coupling constants}
\\
\vskip3mm

\begin{tabular}{|c|c|c|}  \hline
vertex & $g(\GeV)$  & $g^{exp}(\GeV)$ \\
\hline
$\sigma\pi^+\pi^-$ & $3.1\pm0.5$ &$2.6\pm0.2$ \\
\hline
$\kappa K^+\pi^-$ & $3.6\pm0.3$ &$4.5\pm0.4$ \\
\hline
$f_0 K^+K^-$ & $1.6\pm0.1$ & \\
\hline
$f_0 \pi^+\pi^-$ & $0.47\pm0.05$ & $1.6\pm0.8$ \\
\hline
\end{tabular}\end{center}

From Table I we see that, although not exactly 
in between the experimental error bars, the hadronic couplings determined
from the QCD sum rule calculation are
consistent with existing experimental data. The biggest discrepancy is for
$g_{f_0\pi^+\pi^-}$ and this can be understood since the $f_0\to\pi^+\pi^-$ 
decay is mediated by gluon exchange and, therefore,
probably in this case $\alpha_s$ corrections, which were not considered, could
 play an important role.

Here, we follow ref.~\cite{sca} and work at the pion pole. The main reason
for working at the pion pole is that the matrix element in Eq.(\ref{coup})
defines the coupling constant only at the pion pole. For $q^2\neq0$ one would 
have to replace $g_{D_{sJ}D_s\pi}$, in Eq.(\ref{coup}), 
by the form factor $g_{D_{sJ}D_s\pi}(q^2)$ and, therefore, one would 
have to deal with the 
complications associated with the extrapolation of the form factor 
\cite{bclnn,dfnn}. The pion pole method consists in neglecting the pion 
mass in the denominator of Eq.~(\ref{phen}) and working at $q^2=0$. In the 
OPE side one singles out the leading terms in the operator product expansion 
of Eq.(\ref{3po}) that match the $1/q^2$ term. In the phenomenological side, 
in the structure ${q_\mu\over q^2}$ we get:
\beq
\Ga^{phen}(p^2,{\pli}^2)={\la_S m_{D_s}^2f_{D_s}F_{\pi}~
g_{D_{sJ}D_s\pi}\over(m_c+m_s)(p^2-m_{D_{sJ}}^2)({\pli}^2-m_{D_s}^2)}
+\int_{m_c^2}^\infty{\rho_{cont}(p^2,u)\over u-{\pli}^2}~du.
\label{mco}
\enq
In Eq.(\ref{mco}), $\rho_{cont}(p^2,u)$, gives the
continuum contributions, which can be 
parametrized as $\rho_{cont}(p^2,u)={b(u)\over s_0-p^2}\Theta(u-u_0)$ 
\cite{ennr,io2}, with $s_0$ and $u_0$ 
being the continuum thresholds for $D_{sJ}$ and $D_s$ respectively.
Since we are working at $q^2=0$, we take the limit $p^2={\pli}^2$ and we 
apply the Borel	transformation to $p^2\rightarrow M^2$ to get: 
\beq
\Ga^{phen}(M^2)= {\la_S m_{D_s}^2f_{D_s}F_{\pi}~
g_{D_{sJ}D_s\pi}\over(m_c+m_s)(m_{D_{sJ}}^2-m_{D_s}^2)}\left(
e^{-m_{D_s}^2/M^2} -e^{-m_{D_{sJ}}^2/M^2}\right)+A~e^{-s_0/M^2}+
\int_{u_0}^\infty\rho_{cc}(u)~e^{-u/M^2}du,
\label{paco}
\enq
where 
\beq
A=-\int_{u_0}^\infty{b(u)\over(s_0-u)}~du,\;\;\mbox{and }\;\;\rho_{cc}(u)=
{b(u)\over(s_0-u)},
\enq
stands for the pole-continuum transitions and pure continuum contributions.
For simplicity, one assumes that the pure continuum contribution to the 
spectral density, $\rho_{cc}(u)$, is given by the result obtained in the OPE 
side. Asymptotic freedom ensures that equivalence for sufficiently large $u$.
Therefore, one uses the ansatz: $\rho_{cc}(u)=\rho_{OPE}(u)$.
In Eq.(\ref{paco}), $A$ is a parameter which, together with
$g_{D_{sJ}D_s\pi}$, have to be determined by the sum rule.

In the OPE side we work at leading order and  deal with the strange quark as 
a light one and consider the diagrams up to order $m_s$. To keep the charm 
quark mass finite, we
use the momentum-space expression for the charm quark propagator. We calculate
 the light quark part of the correlation
function in the coordinate-space, which is then Fourier transformed to the
momentum space in $D$ dimensions. The resulting light-quark part is combined 
with the charm-quark part before it is dimensionally regularized at $D=4$.
Singling out the leading terms proportional to
$q_\mu/q^2$, we can write the Borel transform of the correlation function in 
the OPE side in terms of a dispersion relation:
\beq
\Ga^{OPE}(M^2)=\int_{m_c^2}^\infty  \rho_{OPE}(u)~e^{- u/M^2}du\;,
\lb{ope}
\enq
where the spectral density, $\rho_{OPE}$, is given by the imaginary part of 
the correlation function. Transferring the pure continuum contribution to
the OPE side, the sum rule for the coupling constant, up to dimension 7, is 
given by:
\beqa
C~\left(e^{-m_{D_s}^2/M^2} -e^{-m_{D_{sJ}}^2/M^2}\right)+A~e^{-s_0/M^2}=
\ga\qq\left[{1\over2^4\pi^2}\int_{m_c^2}^{u_0}du~e^{-u/M^2}u\left(1-
{m_c^2\over u}\right)^2\right.
\nn\\
\left.+{m_cm_s\over8\pi^2}\int_{m_c^2}^{u_0}du~e^{-u/M^2}\left(1-
{m_c^2\over u}\right)-{m_c\ss\over6}e^{-m_c^2/M^2}+{m_s\ss\over12}
e^{-m_c^2/M^2}\left(1+{m_c^2\over M^2}\right)\right],
\label{sr}
\enqa
with 
\beq
C={\la_S m_{D_s}^2f_{D_s}F_{\pi}
\over(m_c+m_s)(m_{D_{sJ}}^2-m_{D_s}^2)}~g_{D_{sJ}D_s\pi}.
\label{coef}
\enq
In Eq.(\ref{sr}),
$\gamma$ measures the isospin symmetry breaking in the quark condensate:
\beq
\ga={\lag0|\bar{d}d-\bar{u}u|0\rag\over\lag0|\bar{u}u|0\rag},
\label{gam}
\enq
and is the source of nonperturbative isospin violation in the OPE
of the vertex function.
The value of $\ga$ has been estimated in a variety of approaches 
\cite{nari2,gale,jim}, with results varying over almost one order of 
magnitude: $-1\times10^{-2}\leq\ga\leq-2\times10^{-3}$. A more recent 
calculation \cite{kms} gives a bigger (in module) value: 
$\ga=-2.6\times10^{-2}$.

In the numerical analysis of the sum rules, the values used for the meson
masses, quark masses and condensates are: $m_{D_{sJ}}=2.317~\GeV$,
$m_{D_s}=1.968~\GeV$, $m_s=0.13\,\GeV$, $m_c=1.2\,\GeV$, 
$\lag\bar{q}q\rag=\,-(0.23)^3\,\GeV^3$,
$\langle\overline{s}s\rangle\,=0.8\lag\bar{q}q\rag$. For the meson
decay constants we use $F_\pi=\sqrt{2}~93\MeV$ and $f_{D_s}=0.22~\GeV$
(obtained using $u_0=6~\GeV^2$ \cite{beni}).
For the current meson coupling, $\la_S$, defined in Eq.(\ref{fp}) we are going
 to use the result obtained from the two-point function in ref.~\cite{blmnn}.
Considering $2.6\leq\sqrt{s_0}\leq2.8~\GeV$ we get $\la_s=(2.9\pm0.3)\times
10^{-3}~\GeV^5$.

In Fig.~1 we show, through the dots, the right-hand side (RHS) of 
Eq.(\ref{sr}), for $u_0=6~\GeV^2$ and $\ga=-1\times10^{-2}$ \cite{jim}, as a 
function of the Borel mass. We use the same Borel window as defined in 
ref.\cite{blmnn}.

\begin{figure}[h] \label{fig1}
\centerline{\epsfig{figure=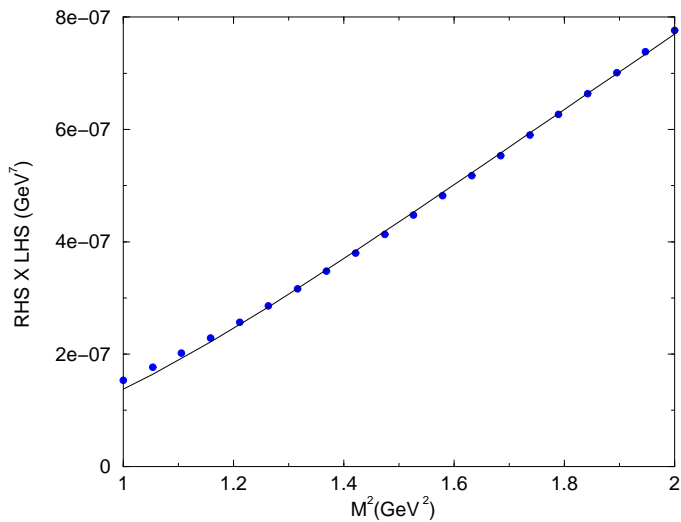,height=70mm}}
\caption{Dots: the RHS of Eq.(\ref{sr}), as a function of the Borel mass.  
The solid line gives the fit of the QCDSR results through 
the LHS of Eq.(\ref{sr}).} 
\end{figure} 

To determine $g_{D_{sJ}D_s\pi}$ we fit the QCDSR results with the analytical
expression in the left-hand side (LHS) of Eq.(\ref{sr}):
\beq
C\left(
e^{-m_{D_s}^2/M^2} -e^{-m_{D_{sJ}}^2/M^2}\right)+A~e^{-s_0/M^2},
\label{exp}
\enq
and we get (using $\sqrt{s_0}=2.7\GeV$): $C=8.34\times10^{-6}~\GeV^7$ and 
$A=6.96\times10^{-6}~\GeV^7$. Using the definition of $C$ in Eq.(\ref{coef})
and $\la_S=2.9\times10^{-3}~\GeV^5$ (the value obtained for $\sqrt{s_0}=2.7
\GeV$) we get $g_{D_{sJ}D_s\pi}=50~\MeV$. Allowing $s_0$ to vary in the
interval $2.6\leq\sqrt{s_0}\leq2.8~\GeV$, the corresponding varyation
obtained for the coupling constant is $45~\MeV\leq g_{D_{sJ}D_s\pi}\leq
55~\MeV$.

Fixing $\sqrt{s_0}=2.7\GeV$ and varying the quark condensate, the charm quark 
and the strange quark masses in the intervals: $-(0.24)^3\leq\lag\bar{q}q\rag
\leq-(0.22)^3\,\GeV^3$, $1.1\leq m_c\leq 1.3\GeV$ and 
$0.11\leq m_s\leq 0.15\GeV$, we get results for the coupling constant
still between the lower and upper limits given above. However, varying
the value of $\ga$ form $-1\times10^{-2}$ to the more recent value given
in \cite{kms}: $\ga=-2.6\times10^{-2}$, and keeping the other parameters 
fixed we get $g_{D_{sJ}D_s\pi}=130~\MeV$. On the other hand, if we use the
smallest (in module) value allowed for $\ga$: $\ga=-2.\times10^{-3}$, we
get $g_{D_{sJ}D_s\pi}=10~\MeV$. Therefore, the biggest source
of uncertainty in our calculation is the value of $\ga$. In all cases
considered here, the quality of the fit between the LHS and the RHS of
Eq.(\ref{sr}) is similar to the one shown in Fig.1.

The coupling constant, $g_{D_{sJ}D_s\pi}$, 
is related with the partial decay width through the relation:
\beqa
\Gamma(D_{sJ}^+(2317)\rightarrow D_s^+\pi^0)=
{1\over 16\pi m_{D_{sJ}}^3}g_{D_{sJ}D_s\pi}^2
\sqrt{\la(m_{D_{sJ}}^2,m_{D_s}^2,m_{\pi}^2)},
\lb{decay}
\enqa
where $\la(a,b,c)=a^2+b^2+c^2-2ab-2ac-2bc$. Considering $\ga=-1\times10^{-2}$,
which was the value found in \cite{jim} to be consistent with the 
neutron-proton mass difference in a QCDSR calculation, and allowing $\langle
\bar{q}q\rangle$, $m_c,~m_s$
and $s_0$ to vary in the ranges discussed above we get:
\beq
\Gamma(D_{sJ}^+(2317)\rightarrow D_s^+\pi^0)=(6\pm2)\keV.
\label{fin}
\enq
However, it is important to state that, if the value for $\ga$ found
in ref.\cite{kms} proves to be correct, then the partial decay width could be 
as large as $\Gamma(D_{sJ}^+(2317)\rightarrow D_s^+\pi^0)\sim40\keV$, in 
agreement with the QCDSR calculation done in ref.\cite{whz}, where the meson
$D_{sJ}^+(2317)$ is considered as a ordinary $c\bar{s}$ state.

In Table II we show the partial decay width obtained by different theoretical 
groups.
The first five calculations assume a $c\bar{s}$ picture for $D_{sJ}^+(2317)$,
while the last two assume a four-quark picture for it.
\begin{center}
\small{{\bf Table II:} The decay width $\Gamma(D_{sJ}^+(2317)\rightarrow 
D_s^+\pi^0)$ from various theoretical approaches.}
\\
\vskip0.3cm
\begin{tabular}{|c|c|c|c|c|c|c|c|}  \hline
ref.& \cite{bar}  &\cite{cf}  &\cite{god2}& \cite{fr}&  \cite{whz}& \cite{ch}
& this work  \\
\hline
$\Ga$ (keV)& $~21.5~$ &$~7\pm1~$  & $~\sim10~$& $~16~$&$~39\pm5~$&$~10-100~$
&$~6\pm2~$\\
\hline
\end{tabular}\end{center}

From the results in Table II we see that we can not get a definitive answer
about the structure of the $D_{sJ}(2317)$ meson from its strong decay width, 
since in both pictures: ordinary $c\bar{s}$ or four-quark states, different 
approachs can give results varying from a few keV to a hundred keV.

We have presented a QCD sum rule study of the vertex function associated with
the strong decay $D_{sJ}^+(2317)\rightarrow D_s^+\pi^0$, where the
$D_{sJ}^+(2317)$ meson was considered as diquark-antidiquark state. We found 
that the source of isospin violation in our calculation is the
parameter $\gamma={\lag0|\bar{d}d-\bar{u}u|0\rag\over\lag0|\bar{u}u|0\rag}$, 
which measures the isospin symmetry breaking in the 
quark condensate. Since, in our approach, the partial decay width is directly 
proportional to $\ga^2$, and since there is a large uncertainty in the value 
of $\ga$, considering  $\ga$ in the range $-2.6\times10^{-2}\leq\ga\leq-2
\times10^{-3}$ we get the partial decay width in the range $0.2\keV\leq
\Gamma(D_{sJ}^+
(2317)\rightarrow D_s^+\pi^0)\leq40\keV$. However, from other QCDSR 
calculation, we believe that the value of $\ga$ is $\sim-1\times10^{-2}$,
which gives the result shown in Eq.(\ref{fin}).

As a final remark we would like to point out that if, instead of using a
isoscalar current, we have used a isovector current for $D_{sJ}(2317)$
(as suggested in ref.~\cite{hte}), the
difference in Eq.(\ref{sr}) would be a factor 2 in the place of $\ga$. In this
case the decay would be isospin allowed and the partial width would be 
$\sim230~\MeV$, much bigger than the experimental upper limit to the total
width $\Gamma\sim5~\MeV$.

\vspace{1cm}
 
\underline{Acknowledgements}: 
I would like to thank F.S. Navarra  and I. Bediaga for fruitful discussions. 
This work has been supported by CNPq and FAPESP.

\end{document}